# Vibration control in plates by uniformly distributed PZT actuators interconnected via electric networks

Stefano Vidoli, Francesco dell'Isola *

*Dipartimento di Ingegneria Strutturale e Geotecnica, Via Eudossiana 18, 00184 Roma, Italy*

**Abstract** – In this paper a novel device aimed at controlling the mechanical vibrations of plates by means of a set of electrically-interconnected piezoelectric actuators is described. The actuators are embedded uniformly in the plate wherein they connect every node of an electric network to ground, thus playing the two-fold role of capacitive element in the electric network and of couple suppliers. A mathematical model is introduced to describe the propagation of electro-mechanical waves in the device; its validity is restricted to the case of wave-forms with wave-length greater than the dimension of the piezoelectric actuators used. A self-resonance criterion is established which assures the possibility of electro-mechanical energy exchange. Finally the problem of vibration control in simply supported and clamped plates is addressed; the optimal net-impedance is determined. The results indicate that the proposed device can improve the performances of piezoelectric actuation.

**internal resonance / equivalent circuits**

## 1. Introduction

In (dell'Isola and Vidoli, 1998-a) and (dell'Isola and Vidoli, 1998-b) the problem of controlling a certain class of truss modular beams has been addressed; it is shown that the available piezoelectric actuators are in principle able to dampen the mechanical vibrations of the beams. In (Vidoli and dell'Isola, 2000) the concept of continuously distributed control has been introduced and developed for one-dimensional beams. This control is obtained by (i) embedding the actuators in the structural members, (ii) interconnecting them by an electric transmission line.

In the present paper it is proposed to control similarly the vibrations in plates by means of a set of uniformly-distributed electrically-interconnected actuators. However, because the plate is a two-dimensional structural member, such an interconnection must be obtained by means of an electrical two-dimensional uniformly-distributed network. The mathematical difficulty (in comparison with (Vidoli and dell'Isola, 2000)) to be confronted concerns the need of changing the kinematical descriptors of the electric state of the system. This difficulty goes along with the need to conceive a way, which is intrinsically two-dimensional, to suitably interconnect the actuators. Both these difficulties are resolved by: (i) introducing the field of e-state (the material time derivative of which is the electric potential); and (ii) conceiving a truly bidimensional circuital interconnection scheme among the actuators which is governed by the same Partial Differential Equation (PDE) valid for a membrane.

We utilize the classical idea of using actuators for achieving damping. However, to the best of our knowledge, nowhere in the literature [1] has a system constituted by: (i) a plate; (ii) an embedded uniformly-distributed set of actuators; (iii) a transmission bidimensional electric network interconneting the actuators, been considered. In the following sections we show that it is possible to find a particular circuital topology which allows for efficient damping of mechanical vibrations. This damping effect is obtained using inductances lower than those needed in the shunt devices proposed up till now (see (Guran and Inman, 1995)). This engineering concept is novel and may be useful in many applications.

*1.1. Advantages of proposed uniformly distributed net-control systems*

The common features of the control devices already conceived are the differentiation between the sensing and the actuation systems and the localization of PZT actuators in a small number of sites of the vibrating structure. Both limit control efficiency; indeed, the first feature implies the need of a coordinating active system, controlling the actions of the actuators in response to the inputs from the sensors. On the other hand the latter feature implies an optimal localization problem – for both actuators and sensors – the solutions of which depend on the particular mechanical vibration mode to be considered (see (Fuller et al., 1996)). Moreover it is difficult to optimize the characteristics of the control system to obtain low equivalent impedances – these are required to allow for a relevant energy transformation from the mechanical to the electrical form – and efficiently drive the PZT actuators (see (Hagood and Von Flotov, 1991) and (Bondoux, 1996)).

Some efforts to overcome the first of these drawbacks have been made. In particular, the concept of self-sensing actuators has been introduced (see (Fuller et al., 1996)); an ad hoc electric circuit is connected to the piezoelectric patch allowing its two-fold behavior. However every patch remains isolated and its electro-mechanical action has to be coordinated with the rest of the structure. Furthermore even when a large number of actuators is used to control the shape of plates (see (Batra and Ghosh, 1995)), interconnecting the actuators via a circuital network was never considered.

In the present paper exploiting the concept of 'analogy' between mechanical structures and electric control systems, it is proposed to control a plate-like structure by means of a distribution of actuators connected to an electric transmission network. An internal resonance phenomenon, between structural modes and electric modes, is induced to obtain the maximum control efficiency.

The net-control system has two practical advantages: ceteris paribus, it requires lower performances to the PZT actuators and allows for: (i) a more efficient control; and (ii) a reduced time to transfer the energy between electrical and mechanical forms. Moreover the net-control system bypasses the problems of optimal positioning (of actuators and sensors) being able to manage all mechanical modes thought the same distributed configuration of its collocated actuators.

In order to prove the previous statements a mathematical model for the conceived apparatus is used to study its coupled electro-mechanical vibrations. Using numerical simulations for square plates, results of a more general nature are obtained. Indeed the considered case yields important information regarding how to establish the coupling of electrical with mechanical modes. The very nature of the *tuning* of the electrical network to the mechanical propagation phenomena is local and therefore independent of the global geometry of the structural member.

---

[1] During the reviewing process the papers (Hoffmann and Botkin, 1999; Botkin, 1999) and (Hoffmann and Botkin, 2000) were found. These papers deal with piezoelectric continuously distributed actuators embedded in plates and exhibit very interesting and mathematically rigorous homogenization results. It has to be underlined, however, that the device modelled there is different from what is considered here. Indeed, in this paper, the uniformly distributed actuators are interconnected by a resonant passive electrical network.

After choosing a circuital topology, tuning is obtained by suitably determining circuital impedances. This is done by means of standard mathematical tools, however, the electro-mechanical coupling effect under investigation is not studied nor is the energy exchange in infinite-dimensional partitioned dynamical systems explicitly considered in the literature (for a detailed discussion of this point see (Vidoli and dell'Isola, 2000)). Therefore we needed to develop an electro-mechanical coupling criterion and find an optimality condition for mechanical energy dissipation.

For dealing with more complicated geometries, suitable finite element schemes will be needed. However, the simple case presented herein seems to be of both theoretical and practical interest. It will allow for a better understanding of the general properties of the conceived device and therefore will direct the design of piezo-electromechanical plates with more complex geometries and will represent a benchmark for future numerical simulation.

*1.2. Description of the system*

In this paper it is proposed to control the mechanical vibrations of a plate with a distribution of piezoelectric actuators interconnected by means of electric impedances, as shown in *figure 1*: gray boxes represent the actuators while the black ones represent the electric impedances. The concept of electrical interconnection of piezoelectric actuators represents the main feature of the considered device. It has to be stressed that this interconnection can be obtained by means of several different circuital topologies; one possible topology is reproduced in *figu e 2*.[2] The fundamental topological difference between the considered bidimensional network and a transmission line has to be underlined: a node is grounded by means of an actuation device and is connected to four other nodes. This circumstance is accounted for mathematically by introducing a PDE governing the evolution of the electric state descriptor.

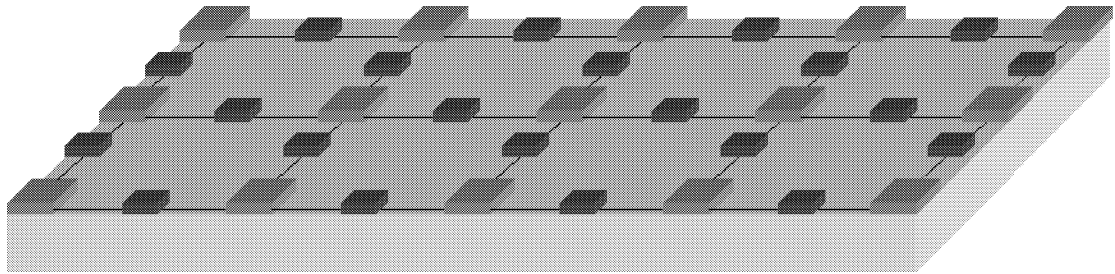

**Figure 1.** Assembled plate and network.

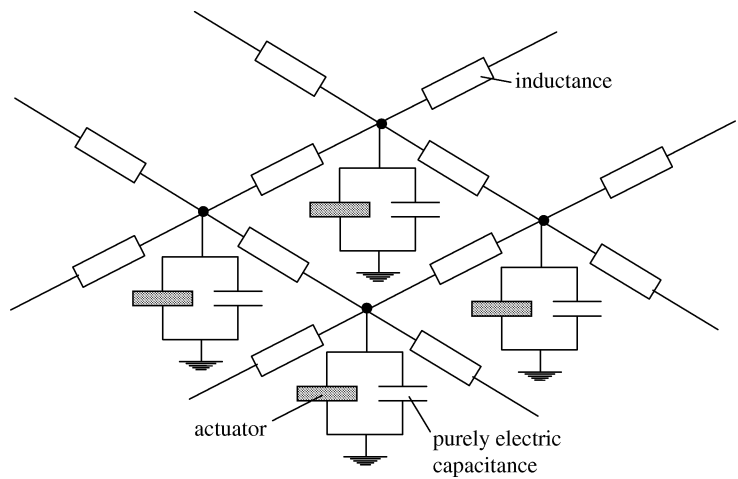

**Figure 2.** Electrical connection scheme.

From an electrical viewpoint the actuators can be considered as capacitors: therefore they are able to store electric energy in DC regimes. A capacitor is a one-port circuital element with two terminals. In the connection scheme represented in *figu e 2* one of these terminals is connected to the electric network while the other is grounded. On the other hand, from a mechanical viewpoint, the actuators: (i) exert mechanical actions in response to electric inputs; (ii) contribute to the structural stiffness of the structure to which they are mechanically connected; and (iii) provide electric signals in response to mechanical deformations.

Therefore the piezoelectric actuators establish an electro-mechanical interaction and allow for the transformation of mechanical into electrical energy. On the other hand the interconnecting electric network allows the electric signal and electric energy to travel simultaneously with the mechanical waves.

The device described in this subsection will be called a *piezo-electromechanical plate*.

## 2. Mathematical model

To start the study of the behavior of a piezo-electromechanical plate, a continuum model is introduced which allows for an accurate description of vibrations when the involved wave-lengths are not too small compared to the dimensions of the single actuator.

The procedure which is followed leads to a homogenized, continuous model that is in some way similar to the homogenized model introduced in (Hoffmann and Botkin, 1999; Botkin, 1999) and (Hoffmann and Botkin, 2000). However we limit ourselves to heuristically deduce it with a formal identif cation in powers, as done usually in the theory of plates (see the classical results found in (Reissner, 1996)) or in a different context e.g. by (Di Carlo et al., 1990; dell'Isola et al., 1998). A rigorous mathematical proof – establishing that the homogenized model we f nd is really a $\Gamma$-limit of the micro-model from which we are starting – is not attempted.

The subsequently postulated balance of power allows us to obtain a set of PDEs for the f elds describing the electric and mechanical states.

In the following formulation, the reader will f nd a multiplicity of mathematical symbols; for instance one will need to distinguish between the Laplacian (respectively gradient) operator in three dimensions and the Laplacian (respectively gradient) operator in two dimensions. This need is justif ed when one considers that the aim is to deduce some equations for a 2-D plate starting from the three-dimensional equations of a Cauchy 3-D continuum. This multiplicity seems to be unavoidable, as it ref ects the complexity of the physical system we try to describe.

We follow the standard mathematical notation; for instance given a second-order tensor $A$ we will denote its symmetric part by sym $A$ and its trace by tr $A$. Given a two-dimensional surface $\mathcal{S}$ embedded in the three-dimensional Euclidean space we will denote by $\partial \mathcal{S}$ its boundary in the sense of Gaussian differential geometry, $\times$ will denote set Cartesian product, $\cdot$ will denote the inner product between vectors, an upper dot will denote the time derivative of considered f elds. Finally $\otimes$ will denote the standard tensor product operation.

---

[2] More complicated and maybe more efficien circuital topologies are conceivable: this being the object of further investigations.

*2.1. Equation of motion*

We deal with a plate body $\mathcal{B}$ occupying the region $\mathcal{C} = \mathcal{S} \times \mathcal{I}$, where $\mathcal{S}$ is a plane surface and $\mathcal{I}$ the real interval $[-h, h]$. As usual, the thickness $2h$ is assumed to be small compared with the diameter of $\mathcal{S}$.

Moreover we assume that the following balance of power:[3]

$$\int_{\mathcal{S}}\int_{\mathcal{I}} (\mathbf{b}\cdot\dot{\mathbf{u}}) + \int_{\partial\mathcal{S}}\int_{\mathcal{I}} (\mathbf{f}\cdot\dot{\mathbf{u}}) + \int_{\mathcal{S}} (i\dot{\psi}) + \int_{\partial\mathcal{S}} (\chi\dot{\psi}) = \int_{\mathcal{S}}\int_{\mathcal{I}} [\mathbf{S}\cdot(\text{sym Grad}\,\dot{\mathbf{u}})] + \int_{\mathcal{S}} (\mathbf{I}\cdot\text{grad}\,\dot{\psi}), \quad (1)$$

must hold for every test field $(\dot{\mathbf{u}}, \dot{\psi})$. Here $\mathbf{u}$ represents the displacement field, while $\dot{\psi}$ is the electric potential field. As a consequence $\dot{\mathbf{u}}$ is the velocity field and $\psi$ is the time-integral of the electric potential difference between the nodes and the ground. Moreover $\mathbf{b}$ and $\mathbf{f}$, $i$ and $\chi$, Grad and grad are the body and surface external forces, the body and surface current densities from the ground, and the gradient operators in $\mathcal{C}$ and $\mathcal{S}$ respectively. Finally $\mathbf{S}$ is the stress tensor and $\mathbf{I}$ is the current vector in the network.

According to the geometry of the body, the position vector is decomposed as:

$$\mathbf{x} = \mathbf{r} + \zeta\mathbf{e}, \quad (2)$$

where $\mathbf{r}$ is the position vector in $\mathcal{S}$, $\zeta \in \mathcal{I}$ and $\mathbf{e}$ is the unit vector perpendicular to $\mathcal{S}$. To deduce from the 3-dimensional Cauchy model of $\mathcal{B}$ the behavior of a bending plate, we use the Kirchhoff–Love compatible identification procedure based on the following kinematical reduction map for displacements:

$$\mathbf{u}(\mathbf{r}, \zeta) = w(\mathbf{r})\mathbf{e} - \zeta\,\text{grad}\,w(\mathbf{r}). \quad (3)$$

As it is easily checked, the previous equation can be interpreted mechanically stating that every material segment orthogonal to the mean surface of the plate in the reference configuration remains: (i) orthogonal in every considered current configuration; and (ii) has a constant length. The function $w$ models the transverse displacement of the material particles of the plate. It is assumed that the electric potential $\dot{\psi}$ depends on $\mathbf{r}$ only. It follows that the infinitesimal deformation field is expressed as (grad grad denotes the Hessian differential operator):

$$\mathbf{E} = \text{sym}(\text{Grad}\,\mathbf{u}) = -\zeta\,\text{sym}(\text{grad grad}\,w). \quad (4)$$

Substituting in the equation (1) the reduction map (3) we obtain:

$$\int_{\mathcal{S}} (b\dot{w} + \mathbf{B}\cdot\text{grad}\,\dot{w} + i\dot{\psi}) + \int_{\partial\mathcal{S}} (T\dot{w} + \mathbf{m}\cdot\text{grad}\,\dot{w} + \chi\dot{\psi}) = \int_{\mathcal{S}} [\mathbf{M}\cdot\text{sym}(\text{grad grad}\,\dot{w}) + \mathbf{I}\cdot\text{grad}\,\dot{\psi}], \quad (5)$$

where

$$\mathbf{M} = -\int_{\mathcal{I}} \zeta\mathbf{S}|_{\mathcal{S}}, \qquad \mathbf{B} = -\int_{\mathcal{I}} \zeta\mathbf{b}|_{\mathcal{S}}, \qquad b = \int_{\mathcal{I}} \mathbf{b}\cdot\mathbf{e}, \quad (6)$$

$$T = \int_{\mathcal{I}} (\mathbf{f}\cdot\mathbf{e}), \qquad \mathbf{m} = \int_{\mathcal{I}} -\zeta\mathbf{f}|_{\mathcal{S}}, \quad (7)$$

are the dynamical generalized forces in the reduced plate model. Applying the divergence theorem (div denotes the divergence operator applied to the following tensor and $\nu$ is the outward pointing normal vector to $\partial\mathcal{S}$

---

[3] In the following integrals the standard notation used in modern measure theory is used. Therefore for instance the infinitesimal volume $dV$ is not indicated. As the integration domains are indicated no misunderstanding is possible.

belonging to the tangent plane to $\mathcal{S}$), we get:

$$\int_{\mathcal{S}} [(b - \mathrm{div}\,\mathbf{B} - \mathrm{div}\,\mathrm{div}\,\mathbf{M})\dot{w} + (i + \mathrm{div}\,\mathbf{I})\dot{\psi}] \\ + \int_{\partial\mathcal{S}} \{[T + (\mathbf{B} + \mathrm{div}\,\mathbf{M}) \cdot \nu]\dot{w} + (\mathbf{m} - \mathbf{M}\nu) \cdot \mathrm{grad}\,\dot{w} + (\chi - \mathbf{I}\cdot\nu)\dot{\psi}\} = 0 \qquad (8)$$

which must hold for every admissible velocity field $(\dot{w}, \dot{\psi})$; this condition leads to the following balance equations:

$$\mathrm{div}\,\mathrm{div}\,\mathbf{M} + \mathrm{div}\,\mathbf{B} - b = 0, \qquad \mathrm{div}\,\mathbf{I} + i = 0, \quad \mathrm{on}\,\mathcal{S}, \qquad (9)$$

and to the following reduced expression of (8):

$$\int_{\partial\mathcal{S}} \{[T + (\mathbf{B} + \mathrm{div}\,\mathbf{M}) \cdot \nu]\dot{w} + (\mathbf{m} - \mathbf{M}\nu) \cdot \mathrm{grad}\,\dot{w} + (\chi - \mathbf{I}\cdot\nu)\dot{\psi}\} = 0 \qquad (10)$$

from which we can get well-posed boundary conditions.

The equations (9) and (10) represent a reduced model for the piezo-electromechanical plate.

Assuming that the body $\mathcal{B}$ is linear, isotropic and homogeneous (so that we do not need to distinguish in the power balance the actual from the reference configuration) and that the network is linear and dissipative, we get:

$$\mathbf{S} = 2\mu_L \mathbf{E} + \lambda_L (\mathrm{tr}\,\mathbf{E})\mathbf{id}, \qquad \mathbf{b} = -\rho\ddot{\mathbf{u}}, \qquad -\mathrm{grad}\,\dot{\psi} = L_N \dot{\mathbf{I}} + R_N \mathbf{I}, \qquad (11)$$

where $\mu_L$ and $\lambda_L$ are the Lamè moduli, $\rho$ the mass density, $L_N$ and $R_N$ are respectively the net-inductance and net-resistance and $\mathbf{id}$ is the identity operator in $\mathcal{C}$. As a consequence the part of the reduced constitutive equations which does not depend on piezoelectro-mechanical coupling (which we denote $\mathbf{M}_m$ and $i_e$) reads as follows:

$$\mathbf{M}_m = J_{\mathcal{I}}[2\mu_L \,\mathrm{sym}\,(\mathrm{grad}\,\mathrm{grad}\,w) + \lambda_L (\mathrm{lapl}\,w)\mathbf{id}], \qquad i_e = \frac{k_C}{d^2}\ddot{\psi}, \qquad (12)$$

where $J_{\mathcal{I}} = 2h^3/3$, lapl is the Laplacian operator in $\mathcal{S}$ and $k_C$ is the purely electric grounded capacitance, while $d^2$ represents the area of influence of the actuator (namely the area of the plate divided by the number of actuators). For the inertial terms from (11) and (6) we get:

$$\mathbf{B} = -J_{\mathcal{I}}\rho\,\mathrm{grad}\,\ddot{w}, \qquad b = -2h\rho\ddot{w}. \qquad (13)$$

However the piezoelectric actuators have a two-fold behavior: from a mechanical viewpoint they enhance the bending stiffness of the plate and produce bending moments in response to applied voltage; from an electric viewpoint they enhance the grounded capacitance per unit area of the electric net and produce a stored capacitive charge in response to applied curvatures.

Let us introduce an orthonormal coordinate system $(o, \mathbf{e}_1, \mathbf{e}_2)$ in $\mathcal{S}$ wherein the partial derivatives of the field $f$ with respect to the corresponding coordinate variables $x_i$ will be denoted $f_{,i}$. Concerning the part of bending moment tensor induced by the piezoelectric effect $\mathbf{M}_e$ we assume that its component expending power on the component $w_{,12}$ of the curvature vanishes so that the following representation holds:

$$\mathbf{M}_e = M_{11}(\mathbf{e}_1 \otimes \mathbf{e}_1) + M_{22}(\mathbf{e}_2 \otimes \mathbf{e}_2). \qquad (14)$$

In addition we specify the properties of the single PZT actuator used by

$$\left\{\begin{array}{c} M_{11} \\ M_{22} \\ Q/d^2 \end{array}\right\} = \begin{bmatrix} g_{mm} & 0 & -g_{me} \\ 0 & g_{mm} & -g_{me} \\ g_{me} & g_{me} & g_{ee} \end{bmatrix} \left\{\begin{array}{c} w_{,11} \\ w_{,22} \\ \dot{\psi} \end{array}\right\}, \quad (15)$$

where $M_{ii}$ and $w_{,ii}$ are the piezo-electrically induced bending moments and curvatures, while $Q/d^2$ and $\dot{\psi}$ are the charge per unit area and voltage between the actuator plates. The constitutive equation (15) establishes that the actuators can supply electrically induced moments only in two fixed material orthogonal directions and that they cannot supply 'mixed' moments; the orthogonal system introduced accounts for this directionality. Moreover we have assumed that the piezoelectric stiffnesses in $\mathbf{e}_1$ and $\mathbf{e}_2$ are equal; this assumption seems reasonable when using PZT actuators exploiting Poisson effect.

Therefore the overall constitutive relations for $\mathbf{M}$ and $i$ read as follows:

$$\mathbf{M} = \mathbf{M}_m + g_{mm}[w_{,11}(\mathbf{e}_1 \otimes \mathbf{e}_1) + w_{,22}(\mathbf{e}_2 \otimes \mathbf{e}_2)] - g_{me}\dot{\psi}\,\mathbf{id}, \quad (16)$$

$$i = i_e + \frac{\dot{Q}}{d^2} = i_e + g_{ee}\ddot{\psi} + g_{me}\,\mathrm{lapl}\,\dot{w}. \quad (17)$$

Let us now introduce the bending stiffness and capacitance per unit area of the plate:

$$D_P = J_{\mathcal{I}}(2\mu_L + \lambda_L), \qquad C_N := g_{ee} + \frac{k_C}{d^2} =: \frac{k_{ee} + k_C}{d^2}; \quad (18)$$

the balance equations in terms of kinematical fields become:

$$\begin{aligned} D_P\,\mathrm{dlapl}\,w + g_{mm}(w_{,1111} + w_{,2222}) - J_{\mathcal{I}}\rho\,\mathrm{lapl}\,\ddot{w} + 2h\rho\ddot{w} - g_{me}\,\mathrm{lapl}\,\dot{\psi} &= 0, \\ -\,\mathrm{lapl}\,\psi + L_N C_N \ddot{\psi} + R_N C_N \dot{\psi} + R_N g_{me}\,\mathrm{lapl}\,w + L_N g_{me}\,\mathrm{lapl}\,\dot{w} &= 0, \end{aligned} \quad (19)$$

where dlapl denotes the double Laplacian operators in $\mathcal{S}$.

In order to find the dimensionless form of (19) we introduce as spatial characteristic length the diameter of the plate $\ell$, and define $v := w/\ell$, $\phi := \psi/\bar{V}$, the characteristic pulsation [4] $\omega = \frac{\pi}{\ell}\sqrt{\frac{D_P}{M_P}}$ – being $M_P = 2\rho\ell^2 h$ the total mass of the plate – so that:

$$\begin{aligned} \frac{D_P}{2h\rho\ell^4\omega^2}\triangle\triangle v + \frac{g_{mm}}{2h\rho\ell^4\omega^2}(v_{,1111} + v_{,2222}) - \frac{h^2}{3\ell^2}\triangle\ddot{v} + \ddot{v} - \frac{g_{me}\bar{V}}{M_P\ell\omega}\triangle\dot{\phi} &= 0, \\ -\frac{1}{L_N C_N \ell^2\omega^2}\triangle\phi + \ddot{\phi} + \frac{R_N}{L_N\omega}\dot{\phi} + \frac{g_{me}}{C_N\ell\omega\bar{V}}\triangle\dot{v} + \frac{R_N g_{me}}{L_N C_N \bar{V}\ell\omega^2}\triangle v &= 0; \end{aligned} \quad (20)$$

where $\triangle$ is the dimensionless Laplacian operator, $(\ )_{,i}$ and the upper dot denote respectively the dimensionless space and time derivative. Let us introduce the following Sobolev norm:

$$\|f\|_{H^2} = \int f^2 + \int \sum_i (f)_{,i}^2 + \int \sum_{ij}(f)_{,ij}^2; \quad (21)$$

---

[4] This choice implies that the first purely mechanical mode in the case of a simply supported plate has a pulsation equal to $2\pi$.

then we will assume that:

$$\frac{g_{mm}}{2h\rho\ell^4\omega^2} \ll \frac{D_P}{2h\rho\ell^4\omega^2}, \qquad \frac{h^2}{3\ell^2}\|\triangle\ddot{v}\|_{H^2} \ll \|\ddot{v}\|_{H^2}. \tag{22}$$

This last inequalities are valid when the wave-length is much bigger than the plate thickness and when one considers only the lower spatial eigenmodes.

The characteristic e-state parameter $\bar{V} := \sqrt{M_P/C_N}$ is chosen to maintain the symmetry, so that (20) becomes:

$$\begin{aligned}\alpha\triangle\triangle v + \ddot{v} - \gamma\triangle\dot{\phi} &= 0, \\ -\beta\triangle\phi + \ddot{\phi} + \gamma\triangle\dot{v} + \delta\dot{\phi} + \delta\gamma\triangle v &= 0,\end{aligned} \tag{23}$$

where

$$\alpha := \frac{D_P}{M_P\ell^2\omega^2} = \frac{1}{\pi^2}, \qquad \beta := \frac{1}{L_N C_N \ell^2 \omega^2}, \tag{24}$$

$$\gamma := \frac{g_{me}}{\ell\omega}\sqrt{\frac{1}{M_P C_N}}, \qquad \delta := \frac{R_N}{L_N \omega} \tag{25}$$

are dimensionless numbers. We remark that when the electro-mechanical coupling parameter $\gamma$ vanishes, (23) reduces to the uncoupled system of the Kirchhoff–Love plate and membrane-like electric network equations:

$$\begin{aligned}\alpha\triangle\triangle v + \ddot{v} &= 0, \\ \ddot{\phi} + \delta\dot{\phi} &= \beta\triangle\phi.\end{aligned} \tag{26}$$

## 2.2. Partitioned modal analysis

In this section we adapt to the set of equations (23) the reasoning developed in (Vidoli and dell'Isola, 2000).

Let $\mathcal{H}_m$ and $\mathcal{H}_e$ be the subspaces of $\mathbf{L}^2(\mathcal{S})$, the space of $\mathcal{R}$-valued square-integrable functions defined on $\mathcal{S}$ verifying suitable homogeneous boundary and smoothness conditions. Let $v, \dot{v}, \ddot{v} \in \mathcal{H}_m$ and $\phi, \dot{\phi}, \ddot{\phi} \in \mathcal{H}_e$. Let $\mathbf{L}_{mm}$ and $\mathbf{L}_{ee}$ be linear self-adjoint differential operators on $\mathcal{H}_m$ and $\mathcal{H}_e$ respectively, and $\mathbf{G}_{me}^A$ indicate the adjoint of the linear differential operator $\mathbf{G}_{me}$ from $\mathcal{H}_e$ to $\mathcal{H}_m$. We consider the following evolutionary problem:

$$\begin{aligned}\alpha\mathbf{L}_{mm}(v) + \ddot{v} - \gamma\mathbf{G}_{me}(\dot{\phi}) &= 0, \\ \beta\mathbf{L}_{ee}(\phi) + \ddot{\phi} + \gamma\mathbf{G}_{me}^A(\dot{v}) + \delta\dot{\phi} + \delta\gamma\mathbf{G}_{me}^A(v) &= 0,\end{aligned} \tag{27}$$

starting from suitable initial conditions for $v$ and $\phi$. We remark that equations (23) have the structure of (27). The subscripts $m$ and $e$ stand for mechanical and electrical respectively.

In order to study the interaction between the electrical and mechanical components of state descriptors we introduce in $\mathcal{H}_m$ and $\mathcal{H}_e$ the eigenbases supplied by the spectral representations theorem for the self-adjoint operators $\mathbf{L}_{mm}$ and $\mathbf{L}_{ee}$ respectively. Therefore, for every $v \in \mathcal{H}_m, \phi \in \mathcal{H}_e$, we have (Reed and Simon, 1980):

$$v = \sum_h v_h m_h, \qquad \phi = \sum_k \phi_k e_k, \tag{28}$$

$$\mathbf{L}_{mm}(v) = \sum_h \lambda_h v_h m_h, \qquad \mathbf{L}_{ee}(\phi) = \sum_k \nu_k \phi_k e_k. \tag{29}$$

Here $\lambda_h$ and $\nu_k$ respectively denote the eigenvalues of $\mathbf{L}_{mm}$ and $\mathbf{L}_{ee}$, $m_h$ and $e_k$ are the corresponding eigenfunctions, $v_h := \langle v, m_h \rangle_{L^2}$ and $\phi_k := \langle \phi, e_k \rangle_{L^2}$ are the time-dependent Fourier coeffcients. If we defne the scalars:

$$C_{hk} := \langle m_h, \mathbf{G}_{me}(e_k) \rangle_{L^2}, \tag{30}$$

$$C_{kh} := \langle e_k, \mathbf{G}_{me}^A(m_h) \rangle_{L^2} = \langle \mathbf{G}_{me}(e_k), m_h \rangle_{L^2} = C_{hk}, \tag{31}$$

and consider that $m_h$ and $e_k$ are bases of $\mathcal{H}_m$ and $\mathcal{H}_e$ as eigenfunctions of self-adjoint operators, equations (27) can be written:

$$\begin{aligned} \ddot{v}_h + \alpha \lambda_h v_h - \gamma \sum_k C_{hk} \dot{\phi}_k &= 0, \\ \ddot{\phi}_h + \beta \nu_h \phi_h + \delta \dot{\phi}_h + \gamma \sum_k C_{kh} (\delta v_k + \dot{v}_k) &= 0, \end{aligned} \qquad h, k = 1, 2, 3, \ldots. \tag{32}$$

Equations (32) clearly show that the infuence on the mode $m_h$ exerted by the mode $e_k$ is measured by the matrix $C_{hk}$ that we can regard as a modal $e \to m$ coupling matrix. In an absolutely similar way $C_{kh}$ represents the modal $m \to e$ coupling matrix. It is now easy to formulate the following:

CRITERION FOR ELECTRO-MECHANICAL COUPLING. *A necessary condition for the presence of electro-mechanical energy exchange between $e_k$ and $m_h$ modes is:*

$$C_{hk} = \langle m_h, \mathbf{G}_{me}(e_k) \rangle_{L^2} \neq 0. \tag{33}$$

## 3. Results of numerical simulations

In this section we particularize the evolution equations found in the previous section to the case of a square piezo-electromechanical plate, describing with suitable boundary conditions the action of both the mechanical and electrical constraints. The criterion (33) is exploited in order to establish a self-resonance between the mechanical and electrical vibration modes. The results we fnd in the particular case considered are, however, of wider applicability. Indeed the tuning of the electrical impedence to get a relevant energy exchange between the electrical and mechanical modes is obtained in a form which is independent of both the shape of the plate and the considered boundary conditions. The numerical simulations described in the present section show that the proposed concept of electrically coordinated actuation allow for the effcient exploitation of available PZT actuators in vibration suppression of relatively *large* (see *table III*) plates.

### 3.1. Analytical solution for the simply supported square plate

Consider a simply supported square piezo-electromechanical plate of side $\ell$. Assume, moreover, that the electric network interconnecting the PZT actuators – on the boundary – has grounded terminal nodes; in this case the boundary conditions for equations (23) become:

$$v = 0, \qquad \mathbf{M} v = 0, \qquad \phi = 0, \tag{34}$$

on each side of the square domain $\mathcal{S}$. The eigenvalues of the purely mechanical and electrical operators:

$$\mathbf{L}_{mm}(f) = \triangle \triangle f, \qquad \mathbf{L}_{ee}(g) = -\triangle g, \tag{35}$$

**Table I.** Mode labeling.

| $k=1$ | 2 | 3 | 4 | 5 | 6 | 7 | 8 | 9 |
|---|---|---|---|---|---|---|---|---|
| $i_k$ | 1 | 1 | 2 | 2 | 1 | 3 | 2 | 3 | 3 |
| $j_k$ | 1 | 2 | 1 | 2 | 3 | 1 | 3 | 2 | 3 |

are respectively given by:

$$\lambda_k = \pi^4 (i_k^2 + j_k^2)^2, \qquad \nu_k = \pi^2 (i_k^2 + j_k^2). \tag{36}$$

If $\mathbf{r}$ denotes the position of the generic material point of $\mathcal{S}$ in the reference configuration, let us introduce the variables $x_i = (\mathbf{r} \cdot \mathbf{e}_i)/\ell$. Then the eigenfunctions corresponding to the eingenvalues (36) are:

$$e_k = \sin(i_k \pi x_1) \sin(j_k \pi x_2), \qquad m_k = \sin(i_k \pi x_1) \sin(j_k \pi x_2), \tag{37}$$

where the indices $i_k$ and $j_k$ which determine the modal forms are defined according to *table I*.

Recalling that in this case $\mathbf{G}_{me}(g) = \triangle g$, we can compute the matrix $C_{hk}$ getting (summation over repeated indices suppressed):

$$C_{hk} = \int_{\mathcal{S}} \triangle(e_k) m_h = -\pi^2 (i_k^2 + j_k^2) \int_{\mathcal{S}} e_k e_h = -\pi^2 (i_k^2 + j_k^2) \delta_{hk}, \tag{38}$$

where $\delta_{hk}$ denotes the Kronecker delta. Note that, because the eigenfunctions are mutually orthogonal, the coupling matrix is diagonal; thus, the coupling can exists only between corresponding modes!

The system (23) is decomposed into an uncoupled sequence of coupled systems of two equations for two unknown functions like the following:

$$\begin{aligned}\ddot{v}_h + \alpha \lambda_h v_h - \gamma C_{hh} \dot{\phi}_h &= 0, \\ \ddot{\phi}_h + \beta \nu_h \phi_h + \delta \dot{\phi}_h + \gamma C_{hh} (\delta v_h + \dot{v}_h) &= 0, \end{aligned} \qquad h = 1, 2, 3, \ldots, \tag{39}$$

or omitting the subscript $h$:

$$\begin{aligned}\ddot{v} + Av - C\dot{\phi} &= 0, \\ \ddot{\phi} + B\phi + C\dot{v} + D\dot{\phi} + CDv &= 0, \end{aligned} \tag{40}$$

where $A = \alpha \lambda_h$, $B = \beta \nu_h$, $C = \gamma C_{hh}$ and $D = \delta$.

Therefore in the case of a simply supported rectangular plate the membrane-like electric network is able to couple one mechanical mode exactly with one electrical mode so that one can get a self-resonance suitably *tuning* the electric network parameters.

However the parameters appearing in (39) depend on the considered mode number; in general the aforementioned tuning will induce the self-resonance of only one pair of electro-mechanical modes.

Therefore the problem of determining the optimal circuital topology for the interconnection of PZT actuators arises. This problem will be addressed in subsequent investigations. We simply remark here (see also the following *figure 10*) that once a self-resonance between a fixed mechanical and a fixed electrical mode is tuned some weaker (but sometimes non negligible and exploitable for damping purposes) electro-mechanical intermodal energy exchanges can be established.

*3.1.1. Non damped energy exchange*

In order to establish the conditions assuring the maximal energy exchange between the mechanical and electrical states, one could develop the general treatment delineated in the case of one-dimensional electro-mechanical structures in (Vidoli and dell'Isola, 2000). However, with a view towards the applications, we consider here a simplified version of that treatment, studying the 1-1 coupling through equations (40) and then extending the results to multiple couplings.

First of all we analyze the non-dissipative case ($D=0$) of equations (40). Its solution, starting from purely mechanical initial data $v_0$, is the following modulated signal:

$$v(t) = V_1 \cos(\alpha_1 t) + V_2 \cos(\alpha_2 t), \qquad \phi(t) = \Phi_1 \sin(\alpha_1 t) + \Phi_2 \sin(\alpha_2 t), \tag{41}$$

where:

$$\begin{aligned}
\alpha_1 &= \sqrt{\frac{1}{2}\left[(C^2+A+B) - \sqrt{(C^2+A+B)^2 - 4AB}\right]}, \\
\alpha_2 &= \sqrt{\frac{1}{2}\left[(C^2+A+B) + \sqrt{(C^2+A+B)^2 - 4AB}\right]}, \\
V_1 &= \frac{u_0}{2}\left(1 + \frac{C^2-A+B}{\sqrt{(C^2+A+B)^2 - 4AB}}\right), \\
V_2 &= \frac{u_0}{2}\left(1 - \frac{C^2-A+B}{\sqrt{(C^2+A+B)^2 - 4AB}}\right), \\
\Phi_1 &= \frac{\alpha_1^2 - A}{\alpha_1 C} V_1, \qquad \Phi_2 = \frac{\alpha_2^2 - A}{\alpha_2 C} V_2.
\end{aligned} \tag{42}$$

In the hypothesis $C^2 \ll A$, with simple manipulations one can find the low frequency analogical components of the modulated signal i.e.:

$$I_{\text{MAX}} = (V_1 + V_2)\cos\left(\frac{\alpha_1 - \alpha_2}{2}t\right), \qquad I_{\min} = (V_1 - V_2)\sin\left(\frac{\alpha_1 - \alpha_2}{2}t\right), \tag{43}$$

respectively representing the envelopes of the maxima and minima, and, as usual, related to the energy contents.

Since we are interested to the most efficient exchange of energy between the mechanical and electrical forms, we seek values of the parameter $B$ that minimize the amplitude of $I_{\min}$ and the time $T_{\text{tr}}$ elapsed to transform the maximal possible amount of initial energy in electrical form:

$$\min_B |I_{\min}| = \min_B |V_1 - V_2|, \qquad \min_B T_{\text{tr}} = \max_B |\alpha_1 - \alpha_2|. \tag{44}$$

These conditions imply respectively:

$$B_1 = A - C^2, \qquad B_2 = A + C^2, \tag{45}$$

indicating that in the interval $(B_1, B_2)$ we get self-resonance. Everywhere in the following we will assume that the self-resonance condition is $B = A$.

446                                      S. Vidoli, F. dell'Isola

Let us now consider the ratio:

$$\Bbbk = \frac{C^2}{A} = \frac{g_{me}^2}{D_P C_N}, \tag{46}$$

which is much smaller than 1 in the applications considered here. This number plays an important rôle in determining the pulsation of the low-frequency analog components in the chosen self-resonance condition. Indeed the dimensionless time interval $T_{tr}|_{B=A}$ needed to transform the mechanical energy of the considered mode into electrical energy is given as a function of $\Bbbk$ (see *figure 4*) by:

$$T_{tr}|_{B=A} = \frac{1}{2(\sqrt{1+\sqrt{\Bbbk}} - \sqrt{1-\sqrt{\Bbbk}})}. \tag{47}$$

Recall that $T_{tr}$ is a dimensionless time and $T_{tr} = 1$ represents one period of the first purely mechanical mode.

The self-resonance condition $B = A$ also implies:

$$\beta \simeq \frac{\alpha \lambda_h}{\nu_h} = \alpha \pi^2 (i_k^2 + j_k^2). \tag{48}$$

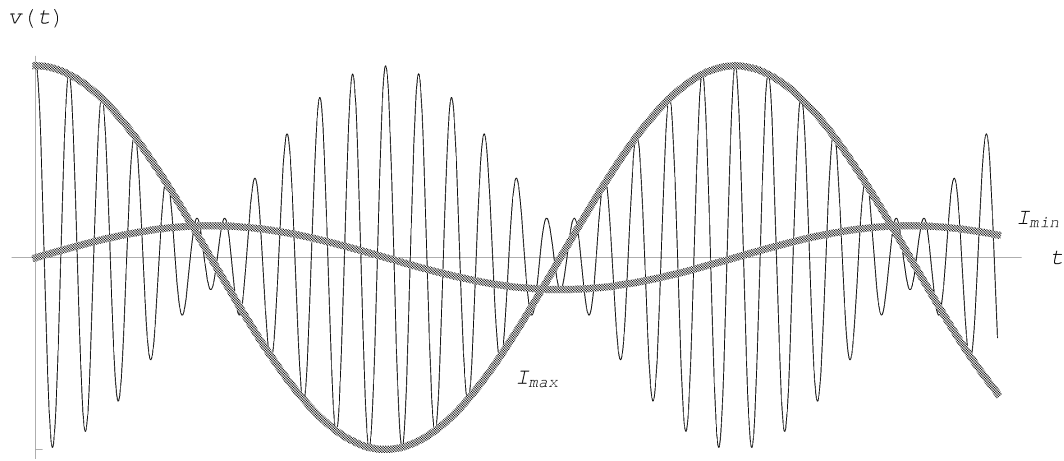

**Figure 3.** Low-frequency analogue components.

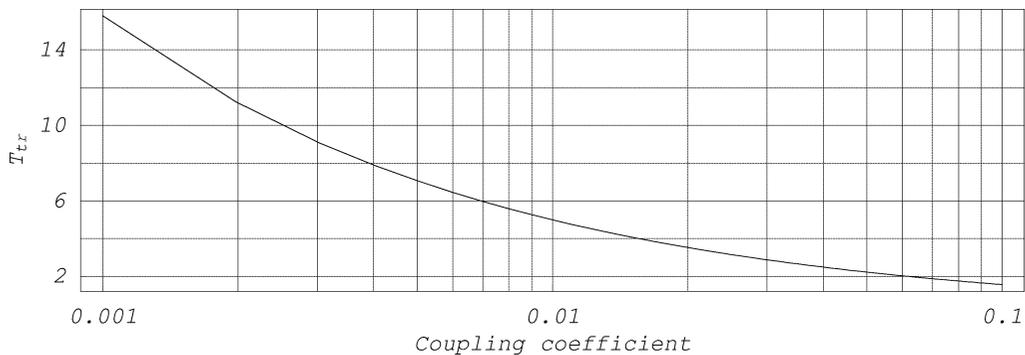

**Figure 4.** Elapsed time to transfer energy $T_{tr}$ as a function of the coupling coefficient $\Bbbk$.

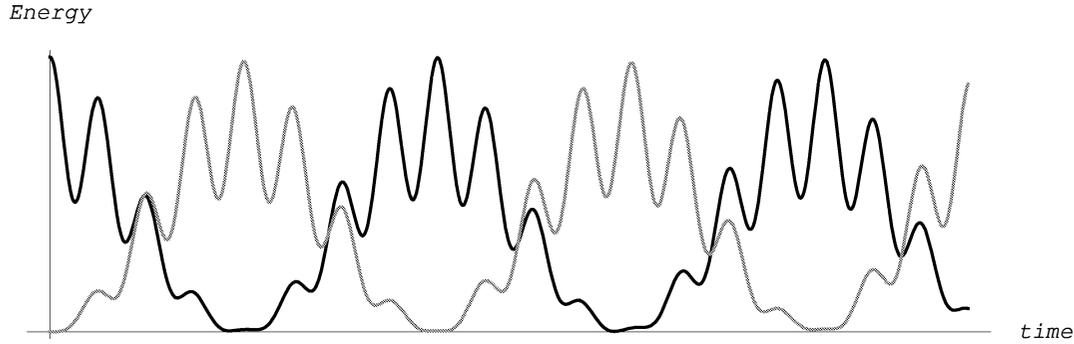

**Figure 5.** Electric (*gray*) and mechanical (*black*) energies vs. time.

Note that $\beta$ depends on the net-inductance $L_N$ that is a tunable parameter. Thus we can simply tune the net-inductance value to couple two modes which, verifying the criterion (33), can be made resonant; the optimal value $L_{Nh}^*$ to couple the $h$-th modes is:

$$L_{Nh}^* = \frac{1}{(i_h^2 + j_h^2) C_N \ell^2 \omega^2} = \frac{1}{(i_h^2 + j_h^2)(k_{ee} + k_C) N_A \omega^2}; \quad (49)$$

here $N_A = \ell^2/d^2$ is the total number of actuators.

The energy flux related to the solution (41) for $B = A$, is visualized in *figure 5* where the thick gray line represents the electric energy.

Note that in *figure 5* there are four different kinds of energies involved:

1. the mechanical elastic energy, $Av^2/2$;
2. the mechanical kinetic energy, $\dot{v}^2/2$;
3. the electric inductive energy, $B\phi^2/2$;
4. the electric capacitive energy, $\dot{\phi}^2/2$,

only the total sum of these energies is constant.

### 3.1.2. Damped energy exchange

The characteristic polynomial of equations (40) is now:

$$P(s) := s^2 C^2 + s D C^2 + (s^2 + A)(s^2 + sD + B) = 0, \quad (50)$$

its complex roots representing the damping ratios (real parts) and the pulsations (imaginary parts) of the associated eigenfunctions.

In *figure 6* a pair of roots (the other pair is the complex conjugate) are drawn as functions of the ratio $D/C$; the gray scale measures the electro-mechanical coupling of the associated eigenvectors: black implies comparable electro-mechanical energy contents. Moreover, the projections of the curves on the planes $\{D/C, -\text{Re}\}$ and $\{D/C, \text{Im}\}$ are drawn.

We observe that:

1. Increasing the ratio $D/C$ (i.e. the net-resistance) definitively leads to the uncoupling of the electro-mechanical wave-forms.

2. The projections on the plane $\{D/C, -\text{Re}\}$ show a maximum of the damping ratio of the root relative to wave-form turning to purely mechanical content when $D/C$ tends to infinity. This circumstance allows for the determination of the critical value for the parameter $D$ proportional to the net-resistance.
3. The projections on the plane $\{D/C, \text{Im}\}$ show that in varying $D$ the eigenfrequencies attain a minimal distance; when $D/C$ tends to infinity the mechanical eigenfrequency tends to $\sqrt{A}$, the electric one vanishes.

A further description of the locus of the roots of the characteristic polynomial (50) can be obtained by introducing the following other polynomial:

$$Q(s) := s^2 C^2 + (s^2 + A)(s^2 + sD + A). \tag{51}$$

As $P$ and $Q$ are real polynomials their roots can be paired by conjugation having coincident (negative because $D$ is positive) real parts. Let us call these real parts $\rho_{Pm}, \rho_{Pe}$ and $\rho_{Qm}, \rho_{Qe}$. It can be proven that:

$$A > 0, \ B > 0, \ D > 0, \ C > 0 \implies (\rho_{Qm}, \rho_{Qe}) \subset (\rho_{Pm}, \rho_{Pe}). \tag{52}$$

Therefore an upper bound for the maximum mechanical damping ratio is obtained by the real parts of the roots of $Q$ when $D = 2C$ and its dimensionless value is $1/2$. These considerations are summarized in *figure 7* where the bold lines represent the real parts of the root of $Q$ and the dashed ones represent the real parts of the roots of $P$ when $B \to A$.

Using the approximated condition $D = 2C$ obtained by means of $Q$ and recalling the definitions of $D$, $C$, $C_{hh}$, and $L_N^*$ we get the following estimated values for the optimal net-resistance:

$$R_N^* = \frac{2\pi^2 g_{me}}{C_N \ell^3 \omega^2} \sqrt{\frac{1}{M_P C_N}} = \frac{2\pi^2 g_{me}}{(k_{ee} + k_C) N_A^{3/2} \omega^2} \sqrt{\frac{1}{M_P (k_{ee} + k_C)}}. \tag{53}$$

Note that $R_N^*$ is, in this case, independent of the mode number; this fact will not hold true when the clamped plate is considered. Remark that both the values of the optimal inductance $L_{Nh}^*$ and resistance $R_N^*$ decrease when a 'more distributed' net-control system is considered (namely when the number of actuators $N_A$ increases).

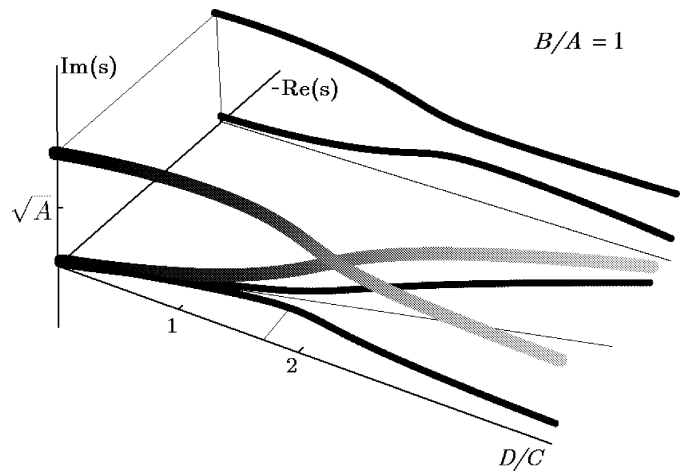

**Figure 6.** Characteristics roots as functions of the ratio $D/C$.

When a dissipative net is considered – in equation (40) $D \neq 0$ –, the energy, once transformed in electric form, is now dissipated and only a fraction transforms back in mechanical form, this is shown in *figure 8*.

In comparing *figure 5* and *8* one should consider that the displayed energy dissipation is obtained for a value of the net-resistance which is not the optimal one.

### 3.2. Numerical solution for the clamped plate

When considering a completely clamped plate a technical difficulty arises in applying the criterion for electro-mechanical coupling. Indeed there is not a treatable closed form for the eigenfunctions of $\mathbf{L}_{mm}$, however one can use their close approximation represented by the product of the eigenfunctions of a clamped–clamped beam (see the first row in *figure 9*); on the other side we still consider a membrane-like network electrically grounded on its boundary so that the eigenfunctions of $\mathbf{L}_{ee}$ will not change (second row in *figure 9*). The labeling of the basis in $\mathbf{L}^2(\mathcal{S})$ we have just chosen trivially parallels that previously introduced in *table I*.

We will apply the coupling criterion to this set of approximated eigenfunctions. The coupling matrix $C$ is no longer diagonal, however, due to the similarities between the modes, the matrix $C$ is quasi-diagonal. *Figure 10* shows a representation of the matrix $C$ by means of the gray scale (blank cell means a vanishing value); the first nine mechanical and the first nine electrical modes are considered.

The novelty in the case of clamped plate is represented by the possibility of coupling one mechanical mode with different electrical ones (if the corresponding $C_{hk}$ is non vanishing). Indeed note that a simple tuning of the net-inductance allows one to make coincident any two frequencies; however all the electric frequencies are shifted together (see *figu e 11*).

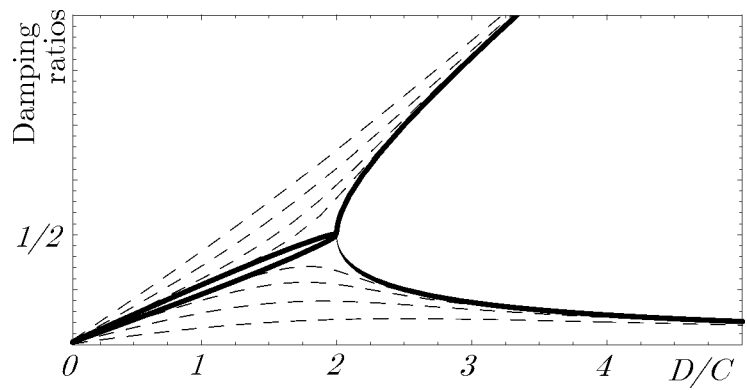

**Figure 7.** Plot of the polynomial $Q(s)$.

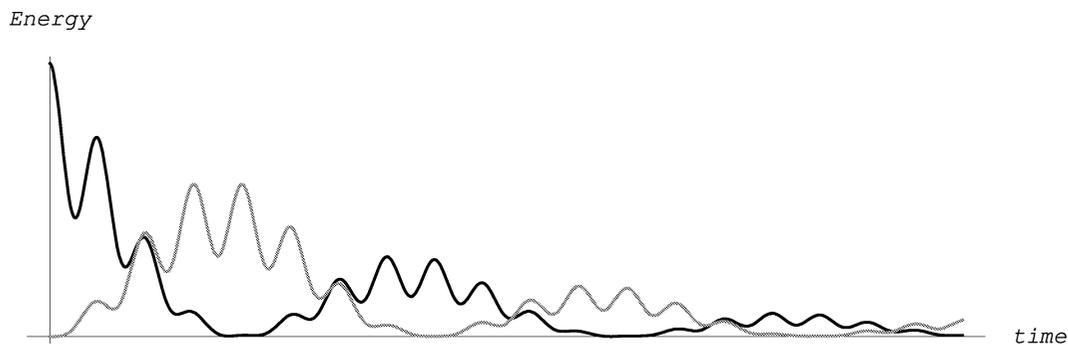

**Figure 8.** Dissipation of electric (*gray*) and mechanical (*black*) energies.

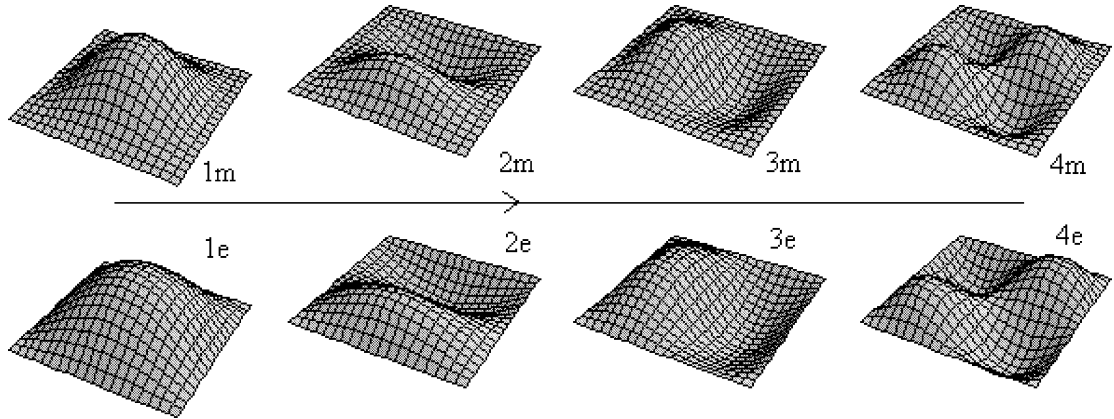

**Figure 9.** Functional basis for the clamped plate.

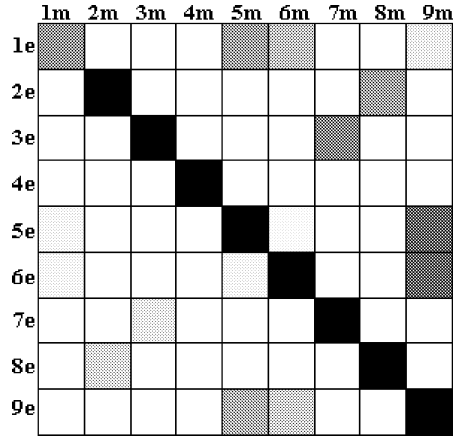

**Figure 10.** Representation of the coupling matrix. A blank cell indicates a vanishing value.

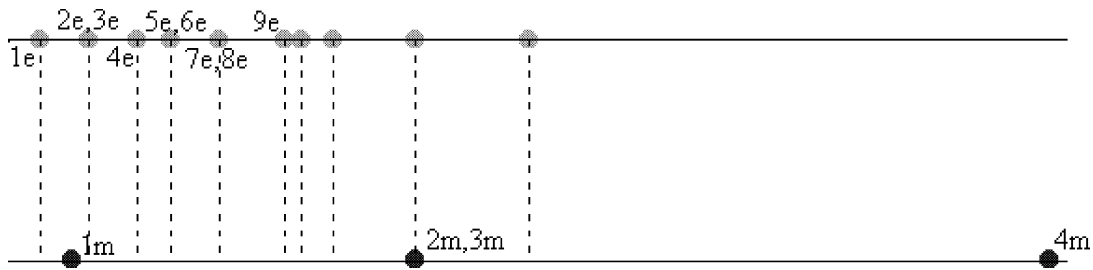

**Figure 11.** Electric and mechanical frequencies.

Table II. Compared eigenvalues.

|  | $k=1$ | 2 | 3 | 4 | 5 | 6 | 7 | 8 | 9 |
|---|---|---|---|---|---|---|---|---|---|
| $\nu_k = (i_k^2 + j_k^2)$ | 2 | 5 | 5 | 8 | 10 | 10 | 13 | 13 | 18 |
| $\lambda_k = (i_k^2 + j_k^2)^2$ | 4 | 25 | 25 | 64 | 100 | 100 | 169 | 169 | 324 |
| $\lambda_k^{(c)}$ | 13.4 | 55.8 | 55.8 | 121.6 | 180.2 | 180.2 | 282.6 | 282.6 | 501.0 |
| $c_k$ | 3.35 | 2.23 | 2.23 | 1.90 | 1.80 | 1.80 | 1.67 | 1.67 | 1.55 |

In order to use the analysis of the previous section in the present case, we list in *table II*:

- the eigenvalues $\nu_k$ of the spatial operator $\mathbf{L}_{ee}(g) = \Delta g$ with $g = 0$ on $\partial \mathcal{S}$;
- the eigenvalues $\lambda_k$ of the spatial operator $\mathbf{L}_{mm}(v) = \Delta\Delta v$ simply supported on $\partial \mathcal{S}$;
- the eigenvalues $\lambda_k^{(c)}$ of the spatial operator $\mathbf{L}_{mm}(v) = \Delta\Delta v$ clamped on $\partial \mathcal{S}$;
- the ratios $c_k = \lambda_k^{(c)}/\lambda_k$.

The eigenvalues $\lambda_k^{(c)}$ have been obtained estimating the Rayleigh ratio on the approximated eigenfunctions. *Table II* is used to determine the optimal inductance and resistance values, namely:

$$L^*_{Nh} = \frac{M_P}{c_h(i_h^2 + j_h^2)C_N \pi^2 D_P}, \qquad R^*_{Nh} = \frac{2g_{me}}{c_h C_N \ell D_P}\sqrt{\frac{M_P}{C_N}}. \tag{54}$$

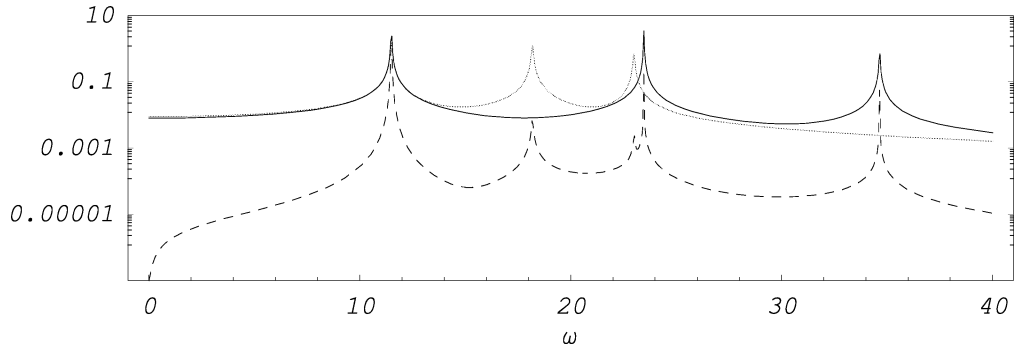

**Figure 12.** Plots of the mechanical (*black*), electrical (*gray*) and electro-mechanical (*dashed*) frequency-response functions. The impedance is set to the optimal value for mode 1.

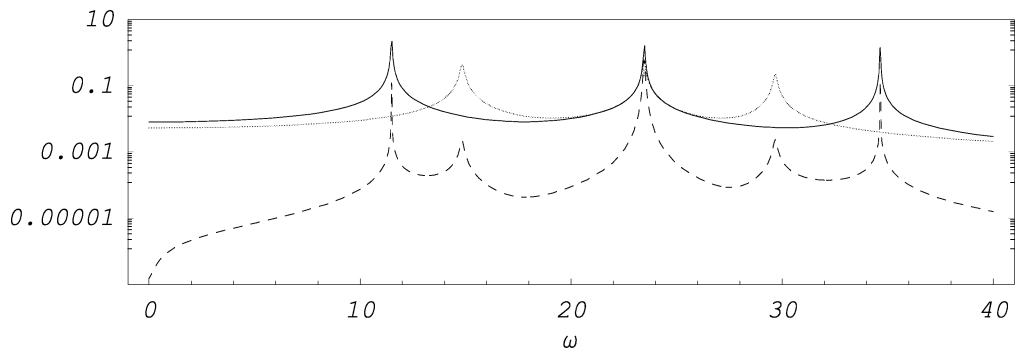

**Figure 13.** Plots of the mechanical (*black*), electrical (*gray*) and electro-mechanical (*dashed*) frequency-response functions. The impedance is set to the optimal value for modes 2 and 3.

**Table III.** Mechanical and electrical parameters.

| | | | |
|---|---|---|---|
| $\ell = 1$ m | $h = 1$ mm | $N_A = 7 \times 7 = 49$ | $g_{me} = 28.10^{-5}$ N V$^{-1}$ |
| $\rho = 2700$ Kg m$^{-3}$ | $E_Y = 70$ GPa | $k_{ee} = 0.6\ \mu$F | $k_C = 1\ \mu$F |

We conclude this section with two plots of the frequency-response functions. In them the net-inductances and resistances are tuned respectively on the optimal values $(L_{N1}^*, R_{N1}^*)$ and $(L_{N2}^*, R_{N2}^*)$. The black and gray lines represent the norms of the purely mechanical and purely electric part of the response function, while the dashed one represents the norm of the coupling part of the response function.

The numerical results, shown in *figures 12* and *13*, confirm the validity of the coupling criterion as a consequence of the optimal impedance values. Indeed the electro-mechanical coupling shows maximal peaks with wider frequency bandwidths when these optimal values are chosen.

In the numerical simulations the values in *table III* were used. The assumed performances of PZT actuators are realistic; they concern the ACX actuators QP20W.

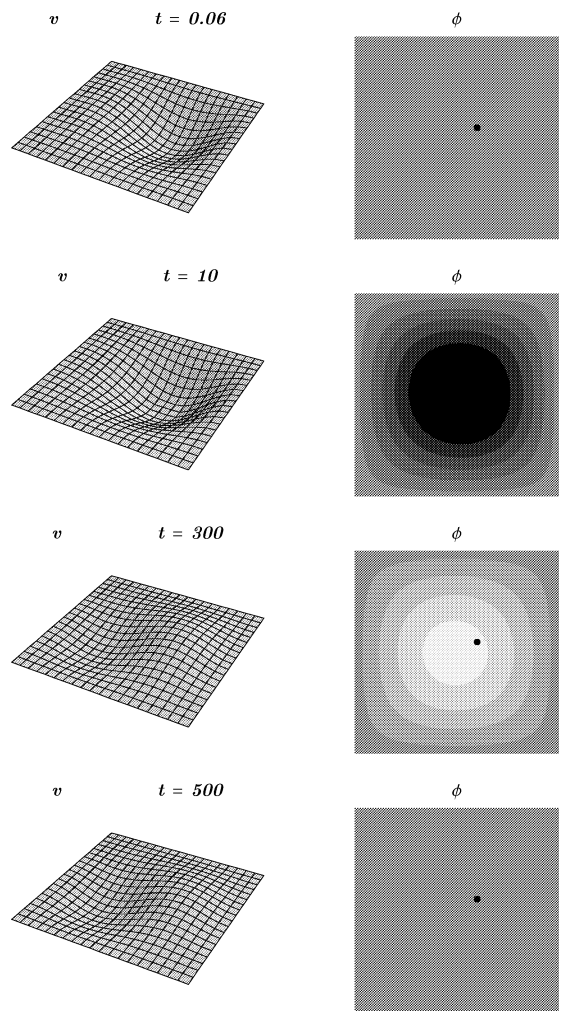

**Figure 14.** Mechanical (left column) and electrical (right column) vibrations induced in the piezo-electromechanical plate by an impulse concentrated at the point $P_1$. Its localization is specified by the black dot. Electric potential is represented as a black–white scale: gray means ground, non-dimensional time is indicated.

*3.3. Response of piezo-electromechanical plates to impulsive external actions*

In this subsection we describe some numerical simulations conceived in order to describe the response of piezo-electromechanical plates to a large class of impulsive external actions. These simulations allowed us to produce some interesting and suggestive animations. The interested reader can download them from the web-site http://www.esm.vt.edu/~henneke/vcpa/animation.

We considered a clamped piezo-electromechanical plate, electrically grounded at its boundary. This last boundary condition means that the electric network interconnecting the PZT actuators behaves as an *edge-supported membrane*.

Moreover it is assumed that only the modes previously labelled $(1, 1)$, $(1, 2)$, $(2, 1)$, $(2, 2)$ (see *table I*) can be considered in Fourier expansion and that initial displacement and electric state are everywhere vanishing. Finally the network is tuned choosing as optimal impedence that found for mode $(1, 1)$.

The impulsive external action is represented by a Dirac $\delta$ initial velocity field concentrated in the two points $P_1$ and $P_2$ of the piezo-electromechanical plate. In the case of vibrations described in *figures 14* and *15* the point $P_1$ has coordinates $(0.6L, 0.55L)$, while those in *figures 16* and *17* are relative to external actions applied in $P_2 = (0.75L, 0.5L)$ where $L$ is the length of the plate side.

The numerical experiment shows that in the first case nearly all electromechanical vibration energy is used to excite the first mode. The black dot represents the localization of $P_1$.

The corresponding time evolution of both mechanical displacement and electric state is shown in *figure 15*, which clearly shows an efficient transfer of energy from mechanical into electrical form.

Rather different is the result represented in the subsequent figures: the impulsive action being concentrated in $P_2$, a non-negligible amount of electromechanical vibration energy is used to excite also the second mode. As the optimal impedences are tuned on the first mode, and as the coupling coefficient between the first and second mode is very small (see *figure 10*), the mechanical vibrations cannot be dissipated using the piezoelectric action of the actuators. This was expected as the electromechanical coupling matrix is nearly diagonal and therefore only a fixed pair of electrical and mechanical modes are coupled.

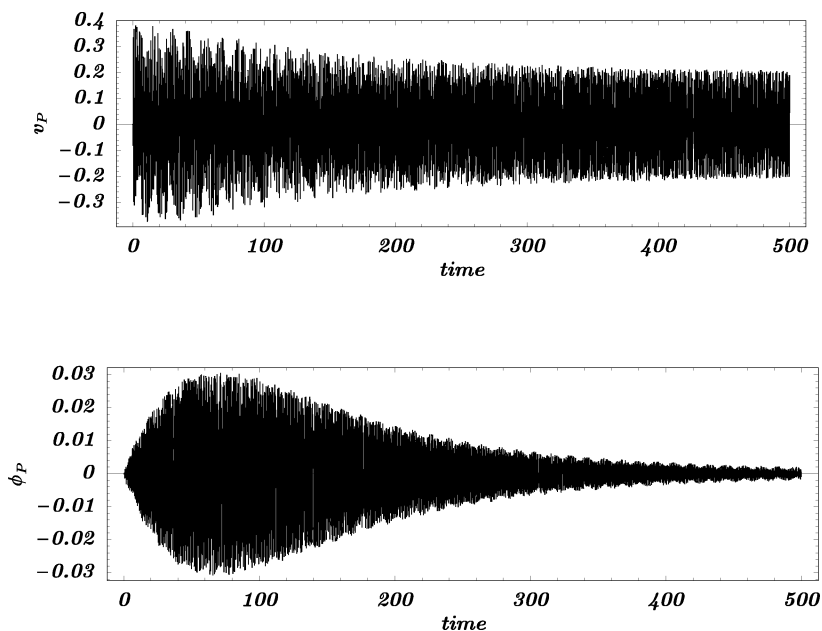

**Figure 15.** Time evolution of displacement and electric state in a representative point of the piezo-electromechanical plate, with impulse concentrated in $P_1$.

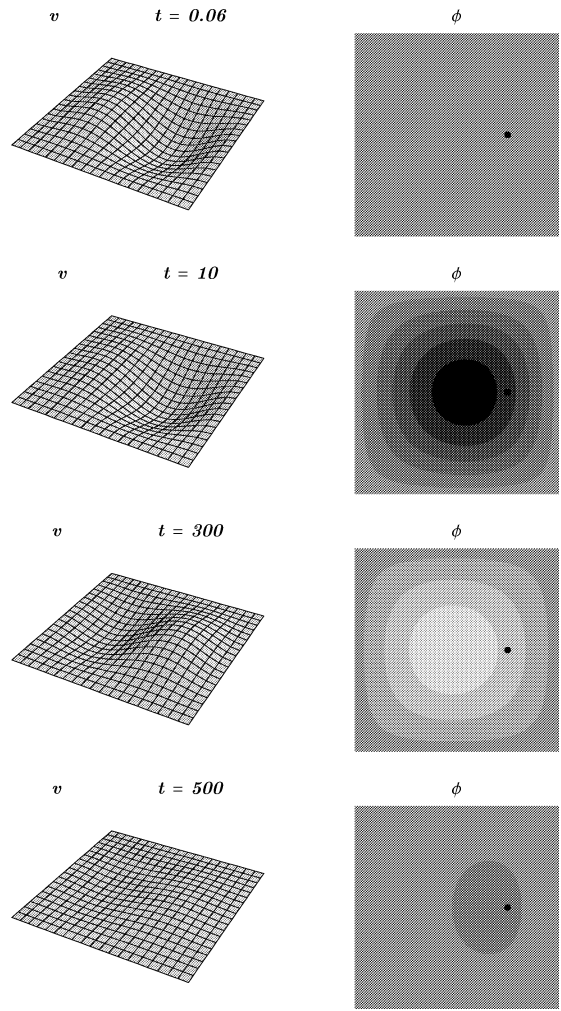

**Figure 16.** Mechanical (left column) and electrical (right column) vibrations induced in the piezo-electromechanical plate by an impulse concentrated at the point $P_2$. Its localization is specified by the black dot. Electric potential is represented as a black–white scale: gray means ground, non-dimensional time is indicated.

The previous considerations are confirmed by the time evolutions plotted in *figure 17*, the mechanical vibration energy is only reduced by the PZT actuation, while after a characteristic time the electric vibration fades. This means that only the mechanical energy initially exciting the first mode is transferred into an electric vibration mode and then dissipated. The non-negligible part of mechanical energy present in the second mechanical mode is not transferred into any electrical mode and is not therefore dissipated.

We conclude that in order to further improve the efficiency of the considered device, more complicated interconnecting electric networks need to be conceived.

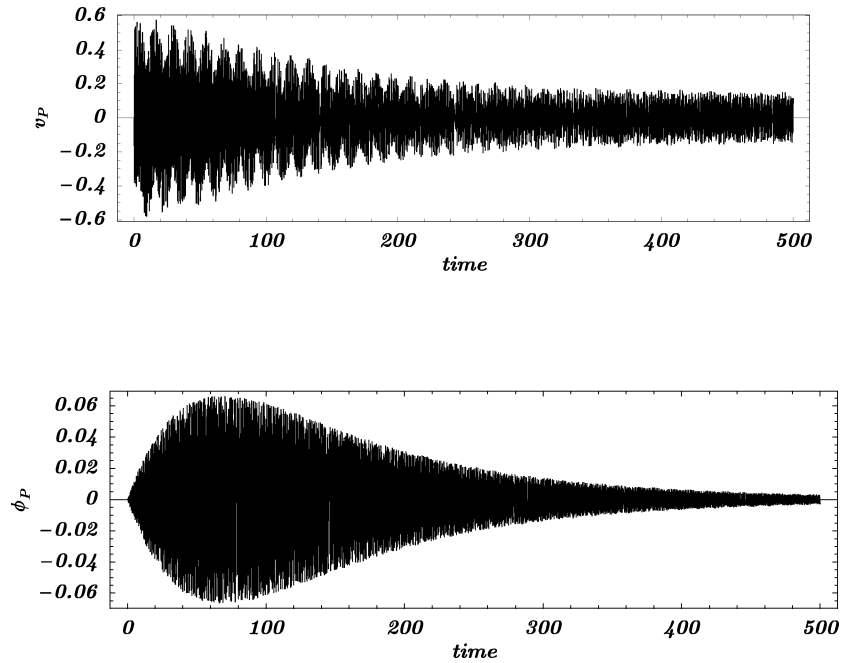

**Figure 17.** Time evolution of displacement and electric state at a representitive point of the piezoelectromechanical plate, with impulse concentrated at $P_2$.

## 4. Conclusions

The device proposed in this paper is based on the concept of global synergic response of a set of PZT actuators to a given mechanical vibration mode. This global response is obtained by conceiving an electric network interconnecting the single actuators. The single electric signal produced by one of them is the potential drop between a node of the electric net and the ground. Thus, PZT actuation is accompanied by an electric waveform which evolves together with the mechanical one. In order to study the performances of the conceived system a mathematical model of its dynamic behavior has been developed. The model is obtained by means of a homogenization procedure and therefore it gives only rough predictions when short wavelengths of the electro-mechanical signals are considered. In the framework of this model, it is proven that a criterion exists assuring electro-mechanical coupling. This criterion allows for the determination of the net-impedances maximizing the electromechanical energy exchange. The eff ciency of the device is indicated by the very low damping ratios which it shows when the optimal net impedance is chosen. This is its main advantage when compared with the devices based on the concept of concentrated actuation (see (Guran and Inman, 1995) and (Fuller et al., 1996)). Furthermore a remarkable decrease of the needed impedance for electro-mechanical coupling is obtained.

At the present stage the construction of a prototype to prove that the concept proposed can be realized in practice, – standing the available technological state of the art – is needed. Two experimental set-ups are being developed at Virginia Polytechnic Institute and State University and Università di Roma *La Sapienza*.

### Acknowledgments

This research was partly performed when the authors were visiting the Department of Engineering Sciences and Mechanics, Virginia Polytechnic Institute and State University, the warm hospitality was greatly appreciated. They gratefully acknowledge Professor E. Henneke, Head of ESM department, and Professor R. Batra for having secured visiting grants. We need to thank Mark D. Carrara M.S. for his help in revising the English form of the paper.